\documentstyle[prb,preprint,aps]{revtex}

\tighten

\begin{document}

\draft

\preprint{LA-UR-94-2354 {\bf /} MA/UC3M/08/94}

\title{Excitation decay in one-dimensional disordered systems with
paired traps}

\author{Angel S\'{a}nchez}
\address{Theoretical Division and Center for Nonlinear Studies, Los
Alamos National Laboratory,\\ Los Alamos, New Mexico 87545\\ and\\
Escuela Polit\'{e}cnica Superior, Universidad Carlos
III de Madrid, c./ Butarque 15,\\ E-28911 Legan\'{e}s,
Madrid, Spain}

\author{Francisco Dom\'{\i}nguez-Adame and Enrique Maci\'{a}$^*$}
\address{Departamento de F\'{\i}sica de Materiales,
Facultad de F\'{\i}sicas, Universidad Complutense,\\
E-28040 Madrid, Spain}

\maketitle

\begin{abstract}

Incoherent transport of excitations in one-dimensional disordered
lattices with pairs of traps placed at random is studied by numerically
solving the corresponding master equation.  Results are compared to the
case of lattices with the same concentration of unpaired traps, and it
is found that pairing of traps causes a slowdown of the decay rate of
both the mean square displacement and the survival probability of
excitations.  We suggest that this result is due to the presence of
larger trap-free segments in the lattices with paired disorder, which
implies that pairing of traps causes less disruption on the dynamics of
excitations.  In the conclusion we discuss the implications of our work,
placing it in a more general context.

\end{abstract}


\narrowtext

\section{Introduction}

Transport properties of randomly disordered systems are a subject of
long-lasting interest both from fundamental and applied
viewpoints.\cite{Alexander,Haus} This issue arises in largely different
physical contexts, many of which can be conveniently mapped onto the
problem of random walks on random lattices.  These include particle or
excitation diffusion in a random one-dimensional (1D) material, low
temperature properties of the random 1D Heisenberg ferromagnet, the 1D
tight binding electron problem with diagonal and off-diagonal disorder,
electrical transmission lines, and excitation transfer along a 1D array
of traps of random depth (see Ref.~\onlinecite{Alexander} and references
therein).  This wide range of applications is the reason why random
walks on random lattices have become a standard model to study transport
in disordered media; in fact, although most of those applications belong
to the field of condensed matter physics, there have been many parallel
pure mathematical and interdisciplinary (biology, chemistry, and
physics) developments.\cite{Haus} On the other hand, in recent years we
have witnessed a great deal of work that tends to undermine well
established beliefs among researchers on the topic of transport in
disordered systems.  In particular, studies on quasiparticle dynamics in
1D systems with correlated disorder
\cite{D,Wu,E,Datta,E2,F,nozotro,D2,PRBKP,PRBF,JPA} have shown that
localization of all eigenstates in 1D disordered systems is not a
general result.  Correlated disorder means that the random parameters of
the system are not independent within a given correlation length.  This
correlation leads to a competition between the short-range order and the
underlying long-range disorder.  Such competition is ultimately the
responsible for the occurrence of unexpected phenomena like, e.g., whole
bands of extended electron states.\cite{PRBKP,JPA} In the few years
elapsed in this decade these results for disordered models exhibiting
non-localization properties have been put on solid grounds.  The
question then arises as to what are the deep physical reasons for this
behavior.

Pursuing further the above line of research, in this paper we concern
ourselves with the study of the decay of {\em incoherent} excitations in
disordered systems, comparing their time decay when correlations are
present to that of purely random systems.  We will use both names, {\em
excitation} and {\em exciton}, to describe our results on quasi-particle
dynamics, since they apply in a more general context.  Note that this
problem is described by a random walk on a random lattice in the way
discussed in the previous paragraph.  We have recently carried out the
time-domain analysis of {\em coherent} (quantum) motion of Frenkel
excitons in 1D systems in the presence of paired correlated traps,
randomly placed in an otherwise perfect lattice.\cite{PRBF} By comparing
with the dynamics of 1D lattices with the same number of unpaired traps,
we found that pairing of traps leads to a slowdown of the survival
probability due to the occurrence of larger segments of the lattice
which are free of traps.  This fact is experimentally relevant since, as
we have argued recently,\cite{PRBO} correlated disorder causes the
occurrence of characteristic lines in the optical spectra of these
systems which are not shared by uncorrelated disorder.  Furthermore,
Scher {\em et al.\/} have shown how for short and intermediate times 1D
transport may be relevant even for three-dimensional systems.\cite{new2}
Then it becomes interesting to elucidate whether this enhacement of the
survival probability due to pairing of traps is restricted to the
coherent motion or, on the contrary, it can be also expected in
incoherent motion of excitons.  This question could be also phrased in
terms of the increase of the survival probability being a quantum effect
or being a general one.  This is the motivation of this work; we report
on it in this paper according to the following scheme.  In Sec.~II we
describe our model and the quantities we are going to use to
characterize it.  In Sec.~III we present our results on survival
probability, mean square displacement, and long-decay asymptotics, and
discuss how they can be interpreted.  We will present numerical
simulations that clearly indicate that pairing of traps leads to a
slowdown of the time decay of incoherent excitations in 1D random
systems.  Also, pairing dramatically affect the excitation size,
measured by its mean square displacement, as a function of time.  Thus,
the main conclusion of the above mentioned calculations is that
structural correlations cause less disruption of the lattice, so the
quasiparticle dynamics is less affected than it would be expected for
the same concentration of traps in a purely random system.  Section IV
concludes the paper with a brief summary of our work and some comments
on implications of its results, which may be of interest in a more
general physical context.

\section{Model}

We consider a 1D lattice whose time evolution is described by the
following master equation for the probability $P_k(t)$ to find the
exciton at site $k$
\begin{equation}
{d\over dt}P_k = F(P_{k+1}+P_{k-1}-2P_k)-G_kP_k,
\label{master}
\end{equation}
where $F>0$ is the intersite rate constant, which will be assumed to be
independent of $k$ hereafter.  Although we restrict ourselves to zero
temperature, thermal effects can be easily included choosing intersite
rate constants depending on temperature according to the Boltzmann
distribution.\cite{Alexander} Here $G_k=G$ if there is a trap at site
$k$ and otherwise $G_k=0$, where $G>0$ is the trapping rate.  Such a
master equation is quite close to those studied in
Refs.~\onlinecite{Alexander,Haus} as general random trapping models.
These have been used as simple theoretical approaches to discuss the
time-dependent effect in fluorescent line-narrowing experiments
concerned with investigations of spectral transfer within
inhomogeneously broadened optical lines (see, for instance,
Ref.~\onlinecite{new}).

The magnitude of interest in luminescence experiments is the survival
probability $n(t)$ defined as
\begin{equation}
n(t)=\langle \sum_k\> P_k(t) \rangle,
\label{survival}
\end{equation}
where the index $k$ runs over all lattice sites and $\langle \ldots
\rangle$ means ensemble average over all possible arrangements of traps.
Moreover, assuming that the excitation is initially at site $k_0$
($P_k(0)=\delta_{kk_0}$), we can also calculate the mean square
displacement of the excitation as follows
\begin{equation}
R^2(t)=\langle \sum_k\> (k-k_0)^2 P_k(t) \rangle,
\label{square}
\end{equation}
where the lattice spacing is taken to be unity hereafter.  These two
functions characterize the exciton dynamics in the lattice.  For
instance, in the absence of traps ($G=0$) it can be shown that $n(t)=1$
and $R^2(t)=2Dt$ in infinite lattices, $D$ being the difussion
coefficient.\cite{Bartolo} We have used those results to test the
reliability of our numerical calculations.  We note that our choice for
the initial condition corresponds to an optical pulsed excitation
experiment where a non-equilibrium localized excitation distribution is
created at site $k_0$ at $t=0$; other possible choices are relevant in
different contexts.  Finally, the correlated disorder is introduced as
follows: We suppose that traps are randomly distributed along the
lattice but with the additional constraint that they only appear in
pairs of neighboring sites (and hence the correlation length is roughly
the lattice spacing).  Hereafter, we define the fraction of traps $c$ as
the ratio between the number of sites with a trap associated with it and
the total number $N$ of sites in the lattice.

\section{Numerical results and discussions}

We have numerically solved the master equation (\ref{master}) for
lattices of $N=1\,000$ sites using an implicit (Crank-Nicholson)
integration scheme.\cite{Recipes} In order to avoid recombinations at
free ends, spatial periodic boundary conditions are introduced.  The
initial condition is, as mentioned before, $P_k(0)=\delta_{kk_0}$, with
$k_0=500$.  Trapping rate $G$ will be measured in units of $F$ whereas
time will be expressed in units of $F^{-1}$.  The maximum integration
time and the integration step were $250$ and $5\times 10^{-4}$,
respectively.  Smaller time steps led to similar results.  Since we are
mainly interested in the effects due to the presence of paired traps
rather than in the effects of the different parameters in the incoherent
motion of excitations, we will fix the values of $F$ and $G$, focusing
our attention on the defect concentration $c$.  Thus we have set $F=1$
and $G=0.2$ henceafter as representative values.  The defect
concentration $c$ ranged from $0.1$ up to $0.9$, and for each lattice a
random distribution of paired traps was chosen.  The ensembles comprised
a number of realizations varying from $50$ to $200$ to check the
convergence of the computed mean values.  The convergence was always
satisfactory between all the ensembles.  In what follows the results we
present correspond to $50$ averages.  In addition lattices with unpaired
traps have been studied and compared with lattices containing the same
fraction of paired traps.  This enable us to separate the effects merely
due to incoherent trapping in one dimension from those aspects that
manisfest the peculiarities of the correlation between random traps.

In our computations we have found that $n(t)$ decays faster as the
fraction of traps increases, in both paired and unpaired traps cases, as
shown in Fig.~\ref{fig1}(a) and Fig.~\ref{fig1}(b), respectively.  This
is expected since trapping should reduce the probability of finding the
excitation in any point of the discrete lattice, and this reduction is
obviously increased on increasing the number of centers able to trap.
In the high concentration limit $c\to 1$ it is not difficult to
demonstrate from (\ref{master}) and (\ref{survival}) that trappping is
simply exponential, $n(t)=\exp(-Gt)$, because in this limit the trap
distribution exhibits translational symmetry and equations can be
exactly solved.  It is worth mentioning that such dependence on time
agrees with the coherent potential approximation (CPA), which is known
to be exact in the high concentration limit.\cite{Movaghar} However,
this is not the case for a random distribution of traps ($c<1$), as seen
in Fig.~\ref{fig1}.  The presence of disorder causes a non-exponential
decay of excitations in systems with either paired or unpaired traps.
We discuss the differences between both kinds of spatial distribution of
traps below.

We have found that another important parameter to describe the time
behavior of excitations is the mean square displacement.  Our results
are shown in Fig.~\ref{fig2}(a) for paired traps and Fig.~\ref{fig2}(b)
for unpaired ones.  In all cases it becomes apparent that the time
evolution of $R^2(t)$ arises from the competition between two processes,
namely diffusion (the exciton is transferred from site to site, starting
at $k_0$) and trapping (the exciton decays in time due to trapping).  At
short times, the first mechanism dominates since the exciton is still
close to the initial position and consequently there are small chances
to be trapped.  On increasing time, the probability of trapping also
increases because the exxiton can be found in a larger segment of the
lattice.  This competition explains the occurrence of a well defined
maximum in $R^2(t)$, whose position depends not only on the
concentration of traps but also on the spatial distribution of traps.
Moreover, the fact that $R^2(t)$ is not a linear function of time is a
consequence of the way we have posed the problem, starting from a
nonequilibrium distribution.\cite{Haus} We elaborate further on these
points later on.

Having described the main features of the incoherent exciton dynamics
and decay due to the presence of traps, we now consider the effects of
pairing of traps in comparison to results obtained in 1D lattices with
unpaired traps.  This comparison will be carried out for systems with
the same fraction of traps, so the differences come simply from the
particular distribution of trapping centers in each kind of lattice.
The main result we found is that, in all cases considered, we have
observed that the exciton decay is slower in the presence of paired
traps.  This is illustrated in Fig.~\ref{fig3} for two different values
of $c$, namely $c=0.1$ and $c=0.4$.  This result is similar to what we
found\cite{PRBF} in the case of quantum transport, as we mentioned in
the introduction hence suggesting that the origin of this slowdown is
similar, that is, pairing of traps causes less disruption of the exciton
motion on the lattice because the average length of segments without
traps is larger in this situation.  This similarity leads us to the
following important conclusion: the slowdown is due only to the
particular distribution of traps, whereas quantum effects do not play
any significative role in these new phenomena.

Another possible way to heuristically understand the above facts is the
following: Consider our master Eq.~(\ref{master}) for two sites which
form one of the paired traps we are discussing, say sites $k$ and $k+1$,
and define $P'\equiv P_k+P_{k+1}$, i.e., the probability to be in any of
the two sites.  By using Eq.~(\ref{master}) for sites $k$ and $k+1$ we
can write down the following equation for $P'$:
\begin{equation}
{d\over dt}P'= F(P_{k+2}+P_{k-1}-P')-G P',
\label{master2}
\end{equation}
where we have taken into account that $G_k=G_{k+1}=G$ as both sites
contain traps.  It can be readily seen that Eq.~(\ref{master2}) is
similar to Eq.~(\ref{master}) but we have renormalized the paired trap
sites into a new, single site, with the same trapping rate but smaller
intersite constant (which in fact violates detailed balance).  From this
we learn that the effect of the paired trap is basically as if it were
one trap; however, the fact that the intersite rate {\em out} of the
(renormalized) site is reduced forces the excitation to stay longer in
it thus increasing the (effective) probability of being trapped.  We see
then that the paired trap can not be trivially compared or dealt with as
if it were a single one.  We note that this argument is just an
heuristic one, as the situation is different if we renormalize one site
which belongs to a pair and one which does not, but it can be seen that
eventually (they should have to be renormalized once again as they would
also be a pair in the renormalized equation) the effect of the pair may
be described in the same way we have just argued.  Of course, this
remains just as a plausibility argument, as further theoretical progress
on the basis of this renormalization procedure seems hopeless in view of
the spatial correlation of the disorder.

Concerning the exciton mean square displacement, we have also compared
results in lattices with the same fraction of paired and unpaired traps.
In Fig.~\ref{fig4} we observe that $R^2(t)$ is always larger when traps
are paired, and that the relative difference between both cases
increases with time.  Such differences are also apparent in the maximum
of $R^2(t)$, as the time of reaching this maximum is always larger in
the case of lattices with paired traps.  Since we have assumed that the
long-time behaviour of $R^2(t)$ is mainly due to trapping effects, this
results reinforce our suggestion that the different exciton behavior in
both kind of systems comes mainly from the particular distribution of
random traps.  There is another feature of Fig.~\ref{fig4} that
deserves attention, namely that $R^2(t)$ is very similar in both paired
and unpaired trap systems up to a time around $t\simeq 30$.  This
similar behavior also shows up in Fig.~\ref{fig3} for $n(t)$.  This is
easily understood if we recall the diffusion-trapping competition we
mentioned in connection with the maximum of the mean square
displacement: Excitation transport properties are {\em
diffusion-dominated} in the early stages of the evolution.  Until a
certain time has elapsed, the chances that the excitation has being
trapped are very small, as it has visited very few trapping sites.  It
is only after this transient that the traps start having a marked effect
on the exciton dynamics.  Therefore, only when transport becomes {\em
trapping-dominated} the differences between paired and unpaired lattices
arise.

Finally, let us consider the asymptotic long-time decay law of excitons
in the presence of traps.  This is an interesting problem and several
theoretical and experimental works have been devoted to find the
relaxation law displayed by excitations in 1D systems.  In the case of
incoherent motion, theoretical predictions show that the survival
probability should decay asymptotically as $\sim \exp(-At^{1/3})$ in the
case of low concentration of unpaired traps\cite{Movaghar,Parris},
whereas there are no available results in the case of paired ones.  We
have studied exciton decay at long times for lattices with paired as
well as unpaired traps.  It has to be noticed that our calculations are
not in the asymptotic limit, especially at low concentration of traps,
when $n(t)$ decays very slowly, so that a direct comparison with
analytical results for $t\to\infty$ could be inconsistent.  On the other
hand, the results are in the experimentally regime, since due to
fluorescent decay processes and finite anisotropies, one can actually
observe one dimensional diffusion processes for only a finite time
span.\cite{Alexander} Plotting $\ln |\ln n(t)|$ versus $\ln t$ in the
range from $t=100$ up to $250$ we have confirmed that survival
probability fit strechted exponentials of the form $n(t) \sim \exp(
-At^\alpha)$ in all cases, as shown in Fig.~\ref{fig5}.  The value of
the parameter $\alpha$ is lower in the case of lattices with paired
traps, hence confirming the fact that disorder correlation reduces the
exciton decay rate even at long time.  It is also interesting to mention
that $\alpha$ depends on the concentration of traps, and it increases
with $c$ in the range of time considered.  At low and moderate values of
$c$ it becomes of order of $0.6$-$0.7$ whereas at higher concentrations
is close to unity.  These results should be regarded only as
qualitatively correct since at very long times roundoff errors increase
while the magnitude of $n(t)$ decreases, and hence many averages are
actually needed to accurately compute the values of the exponent, which
is rather time consuming.  It is then clear that a theoretical
description would be very valuable for a complete understanding of our
results.

\section{Conclusions}

The present paper has been devoted to get a more complete and general
comprehension of the quasiparticle dynamics in 1D systems with
correlated disorder, which is extensively being investigated at present.
In particular, we have focused on incoherent exciton transport in 1D
random lattices with a certain number of traps appearing in pairs along
the lattice, and results have compared to those obtained in the case of
unpaired ones.  In light of computations, we have concluded that
incoherent excitons decays slower when pairing is introduced, in a
similar way as we have previously found in the case of (coherent)
Frenkel excitons.\cite{PRBF} We have also seen that the paired nature of
the traps gives indeed rise to new effects which cannot be simply
understood by treating each paired trap separately, because intersite
transfer rates are also affected.  All these phenomena also manifest
themselves in the square mean displacement, which is found to be larger
in the case of paired traps at all times, and in the long time
asymptotics, described by a smaller exponent in the stretched
exponential dependence.  We stress that this differences should be
noticeable through optical measurements, as in the case of quantum
excitations.\cite{PRBO} Indeed, a most interesting result is that the
increasing of the survival probability and the mean square displacement
are not a quantum effect, but rather, something that comes from the fact
that there is spatial correlation between traps.

The previous paragraph summarizes the conclusions that can be extracted
from our calculations regarding the specific application of the model to
exciton transport properties in solids.  In addition, there are some
issues of more general character that may be learned from what we have
reported.  First, in connection with recent work on suppression of
localization (see Ref.~\onlinecite{PRBKP} for a rather exhaustive list of
references as well as a summary of results) we see that the consequences
of correlation are very different due to the largely disparate
characteristics of wave equations versus diffusion equations: Electrons
and classical waves delocalize, whereas effects on excitations described
by random walks are less dramatically exhibited in longer lifetimes.
However, in both cases, and in spite of being very different problems,
correlation has very profound effects; this suggests that the influence
of having non-white disorder as usually assumed may be important in very
many fields.  Another interesting point is related to applications of
random walks on random lattices in condensed matter physics, such as
those discussed in the introduction.  The analytical treatments
available so far rely heavily on sometimes unrealistic assumptions, i.e,
starting from an equilibrium distribution, like the average-T-matrix
approximation, or having uncorrelated distributions of traps, like the
effective medium approximation.\cite{Emilio} To our knowledge, our
results are the first ones on models which verify none of both
hypothesis, and theoretical approaches developed to deal with this
problem (probably combining some renormalization procedure to remove the
correlations or reduce their role followed by a description in the
spirit of effective-medium approaches) will be most likely very useful
in other subjects in condensed matter physics.  As a final remark from
the viewpoint of applications, it can be expected that the calculation
presented here will be of use as a means to discern the local spatial
structure of active centers in solids, in experiments using pulsed
initial excitations.  On the other hand, were our results found to be
experimentally relevant, they may be employed to design devices with
special optical properties.  We hope that the numerical work presented
here stimulates parallel advances on the theoretical and experimental
sides.

\acknowledgments

This work is partially supported by Universidad Complutense through
project PR161/93-4811.  A.\ S.\ is partially supported by DGICyT (Spain)
grant PB92-0248, by MEC (Spain)/{}Fulbright, and by the European Union
Network ERBCHRXCT930413.  Work at Los Alamos is performed under the
auspices of the U.S.\ D.o.E.

\begin{figure}
\caption{Logarithm of the survival probability of excitons as a function
of time for lattices of $N=1000$ sites with (a) paired and (b) unpaired
traps. The fraction of traps is $c=0.2$, $0.4$, $0.6$ and $0.8$ from top
to bottom. Each curve comprises the results of $50$ averages.}
\label{fig1}
\end{figure}

\begin{figure}
\caption{Mean square displacement of excitons as a function of time for
lattices of $N=1000$ sites with (a) paired and (b) unpaired traps.  The
fraction of traps is $c=0.2$, $0.4$, $0.6$ and $0.8$ from top to bottom.
Each curve comprises the results of $50$ averages.}
\label{fig2}
\end{figure}

\begin{figure}
\caption{Logarithm of the survival probability of excitons as a function
of time for lattices of $N=1000$ sites with  paired (solid lines) and
unpaired (dashed lines) traps. The fraction of traps is indicated on
each plot, which comprises the results of $50$ averages.}
\label{fig3}
\end{figure}

\begin{figure}
\caption{Mean square displacement of excitons as a function of time for
lattices of $N=1000$ sites with paired (solid lines) and unpaired
(dashed lines) traps.  The fraction of traps is indicated on each plot,
which comprises the results of $50$ averages.}
\label{fig4}
\end{figure}

\begin{figure}
\caption{$\ln |\ln n(t)|$ as a function of $\ln t$ for lattices of
$N=1000$ sites with paired (solid lines) and unpaired (dashed lines)
traps.  The fraction of traps is indicated on each plot, which comprises
the results of $50$ averages. From top to bottom, the values of the
slopes are $\alpha=0.88$, $0.79$, $0.78$ and $0.61$.}
\label{fig5}
\end{figure}

\end{document}